\documentclass[doublecol]{epl2}
\usepackage{epsfig,graphicx,tabularx}
\usepackage{amsmath}
\usepackage{color,braket}
\usepackage{url}

\newcommand{\be}{\begin{equation}}
\newcommand{\ee}{\end{equation}}
\newcommand{\sm}[1]{\left|#1\right|^2}
\newcommand{\dd}{\mathrm{d}}

\newcommand{\ti}[1]{\tilde{#1}}

\title{Dynamics of highly unbalanced Bose-Bose mixtures: miscible vs immiscible gases}

\author{A. Sartori\inst{1} \and A. Recati\inst{1}}
\shortauthor{A. Sartori and A. Recati}

\institute{                    
  \inst{1} INO-CNR BEC Center and Dipartimento di Fisica, Universit\`a di Trento, 38123 Povo, Italy
}

\abstract{
We study the collective modes of the minority component of a highly unbalanced Bose-Bose mixtures. In the miscible case the minority component feels an effective external potential and we derive an analytical expression for the mode frequencies. The latter is independent of the minority component interaction strength.
In the immiscible case we find that the ground state can be a two-domain walls soliton. Although the mode frequencies are continuous at the transition, their behaviour is very different with respect to the miscible case. The dynamical behaviour of the solitonic structure and the frequency dependence on the inter- and intra-species interaction is numerically studied using coupled Gross-Pitaevskii equations.}

\begin{document}

\maketitle

\section{Introduction}

Mixtures of Bose-Einstein condensates (BECs) show many interesting phenomena, related to their ground states and excitations, that are not present in single component condensates\cite{MixturesHo,MixturesSvid}. The main ingredient of multi-species BECs is the interplay between intra- and inter-species
interactions, which -- together with the possibility of tailoring the trapping potentials -- gives rise to new ground states and to peculiar dynamic processes. The problem is very general being related to phases and stability of interacting superfluids and coupled non-linear equations. 
In particular superfluid current stability in mixtures \cite{SFcurrent} has been started to be addressed experimentally only very recently \cite{Zoran}.

Moreover, thanks to new technological developments, the highly unbalanced case has attracted a lot of attention both on the theoretical and experimental side with a huge number of different realizations and regimes (see, e.g. \cite{impBECBloch,catani} and reference therein). This is quite obvious being the impurity problem almost ubiquitous in physics, especially in condensed matter \cite{Affleck}. In that respect cold atoms are promising candidate to simulate and shed new light on polaron physics  \cite{jakschPol,timmermans,Tempere}. 

\begin{figure}[ht]
	\centering
	\includegraphics[width=7cm]{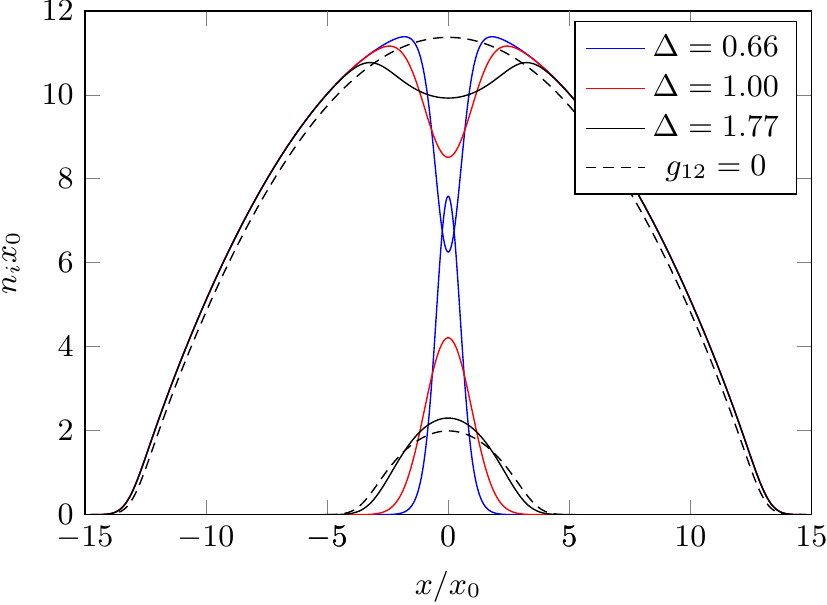}
	\caption{Majority and minority component densities for different values of the parameter $\Delta=g_{11}g_{22}/g_{12}^2$ and for $g_{12}=g_{11}$. The miscible phase is for $\Delta>1$.} \label{fig:DiffD}
\end{figure}

Inspired by the experiment by Catani {\sl et al.}\cite{catani}, in this Letter we are interested in the easiest case of a trapped one-dimensional highly unbalanced mixture with tunable interaction strengths at zero temperature.
We study the ground states and the dynamics of the minority component, depending on whether the gases are miscible or immiscible. In the latter case a novel solitonic solution of the mean-field equations is found. In particular we point out that while looking at the density profiles (see Fig. \ref{fig:DiffD}) the two phases do not differ too much, their dynamical behaviour is very different.  Let us mention here that some interesting results on the breathing mode frequency  for the single-impurity case has been recently obtained in particular regimes by \cite{jaksch}.

The system we consider is a mixture of two one-dimensional Bose gases trapped in harmonic potentials 
\be V_i(x) = \frac{1}{2}m_i\omega_i^2x^2, \quad i=1,2, \ee
where $m_i$ and $\omega_i$ are the mass and the trapping frequency of the $i$-th atomic specie, respectively.

The system at $T=0$  can be described by two coupled time dependent Gross-Pitaevskii equations (GPE) for the condensate wave functions $\Psi_i$\cite{PethickBook}:
\be \label{eq:GPE1}
 i\hbar\frac{\partial}{\partial t} \Psi_1 =\left[-\frac{\hbar^2}{2m_1}\frac{\partial^2}{\partial x^2}+V_1+g_{11}\sm{\Psi_1}+g_{12}\sm{\Psi_2}\right] \Psi_1, \ee
\be \label{eq:GPE2}
i\hbar\frac{\partial}{\partial t} \Psi_2=\left[-\frac{\hbar^2}{2m_2}\frac{\partial^2}{\partial x^2}+V_2+g_{22}\sm{\Psi_2}+g_{12}\sm{\Psi_1}\right] \Psi_2, \ee
where $g_{ij}$, $i,j=1,2$ are the interaction coupling constants. The condensate wave-function are normalized to the number of atoms in each component 
\be N_i=\int\dd x \sm{\Psi_i} \quad i=1,2, \ee
and in the following we will always consider $N_1\gg N_2$.

\section{Miscible Case}

\subsection{Ground State}

It is well known that mixtures can be miscible or immiscible depending on whether the ratio $\Delta=g_{11}g_{22}/g_{12}^2$ is larger or smaller than one. 
We start discussing the miscible case, for which the minority component is embedded in the majority one. 
Within Thomas-Fermi (TF) approximation the two component densities, derived from Eqs.\eqref{eq:GPE1}-\eqref{eq:GPE2}, can be written as \cite{Riboli}
\be 
\label{eq:n2full}
n_2(x) \!\!= \!\!\frac{1}{g_{11}}\frac{\Delta}{\Delta-1} \left[\mu_2-\frac{g_{12}}{g_{11}}\mu_1-V_2+\frac{g_{12}}{g_{11}}V_1\right],n_2(x)>0, 
\ee
\be
n_1(x) = \begin{cases} \frac{1}{g_{11}}\frac{\Delta}{\Delta-1} \left[\mu_1-\frac{g_{12}}{g_{22}}\mu_2-V_1+\frac{g_{12}}{g_{22}}V_2\right], &\!\!\!\mbox{ } n_2(x)>0
\\ \frac{1}{g_{11}}\left[\mu_1-V_1\right], &\!\!\! \mbox{} n_1(x)>0\end{cases}
\ee
where $\mu_i$, $i=1,2$ are the chemical potentials of the two components.

One can get a better insight considering only the first order correction due to the second component. In this case the majority component is not affected by the minority one, while the density of the latter can be put in a standard TF form   
\be
n_2(x)= \frac{1}{g_{22}} \left(\ti{\mu}_2-\frac{1}{2}m_2\ti{\omega}_2^2x^2\right)  \Theta(\tilde{R}_2^2-x^2), \label{eq:n2app}
\ee
by introducing the renormalized chemical potential  $\ti{\mu_2}=\mu_2(1-g_{12}\mu_1/g_{11}\mu_2)$, trapping frequency 
\be\ti{\omega}_2^2=\omega_2^2\left(1-\frac{g_{12}m_1\omega_1^2}{g_{11}m_2\omega_2^2}\right) \label{eq:myresult}
\ee 
and consequently the TF radius $\ti{R}_2^2=2\ti{\mu}_2/m_2\ti{\omega}_2^2$. Which tells us that the minority component in the ground state feels a renormalized external potential due to the mean-field interaction with the majority one. We will see in the next Section that in a certain regime this is the case even for the collective modes.

\subsection{Collective modes}

\begin{figure}[ht]
	\centering
	\includegraphics[width=7cm]{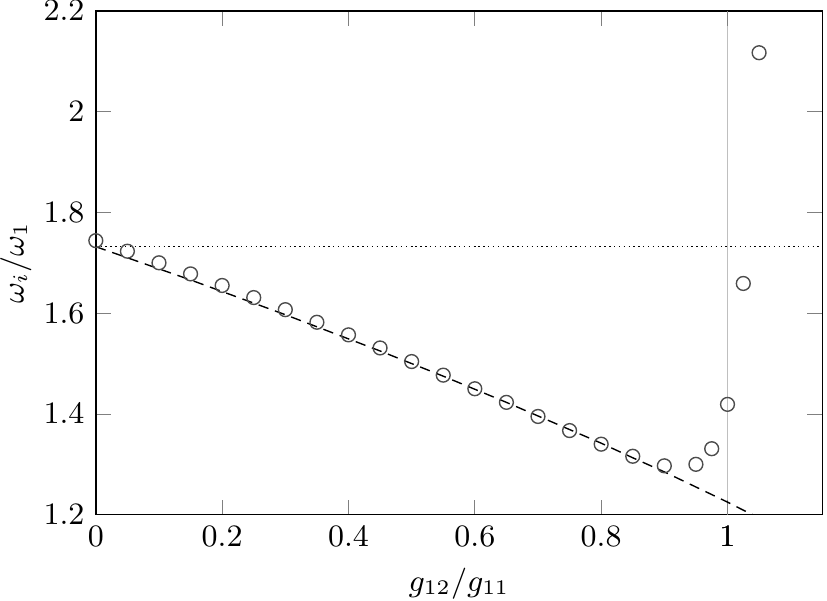}
	\caption{Comparison between numerical and theoretical frequencies for the breathing mode of the minority component. The dashed line is the theoretical prediction, while circles are the numerical results. The parameters are $N_1=200$, $N_2=10$, $m_1=m_2/2=87$~amu, $\omega_1= \omega_2=2\pi\times 200$~Hz and $g_{11}=g_{22}$.} \label{fig:freq-br}
\end{figure}

\begin{figure}[ht]
	\centering
	\includegraphics[width=7cm]{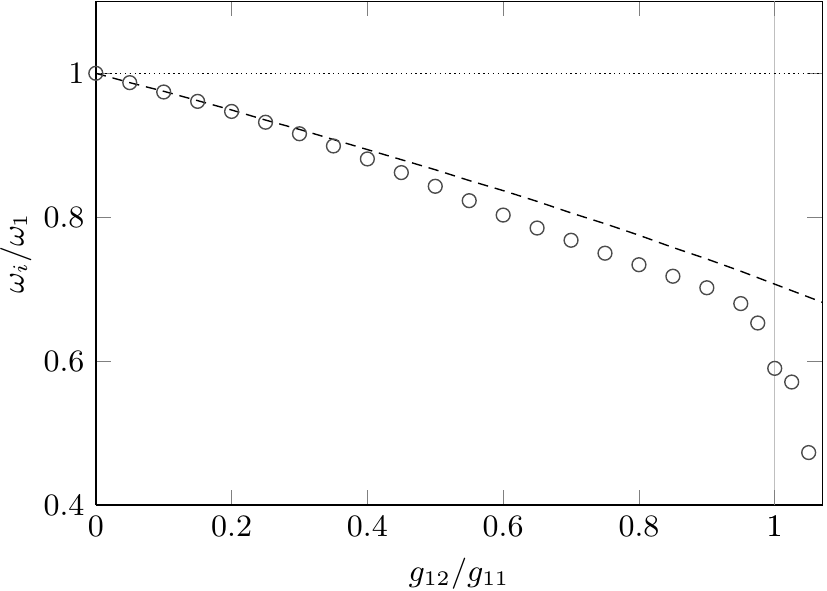}
	\caption{The same as in Fig. \ref{fig:freq-br} but for the center-of-mass mode of the minority component.} \label{fig:freq-cm}
\end{figure}
In order to study analytically the collective oscillations of a BEC we recast the GPEs in the hydrodynamic formulation \cite{sandrolevbook}. Considering the simple approximation where the minority component does not affect the majority one, the hydrodynamic equations read 
\be \label{eq:hy1}
\frac{\partial^2\delta n_1}{\partial t^2} - \frac{\partial}{\partial x}\left(c_1^2(x)\frac{\partial \delta n_1}{\partial x}\right) = 0, \ee
\be \label{eq:hy2}
\frac{\partial^2\delta n_2}{\partial t^2} - \frac{\partial}{\partial x}\left(c_2^2(x)\frac{\partial \delta n_2}{\partial x}\right) = \frac{g_{12}}{g_{11}} \frac{\partial}{\partial x}\left(c_2^2(x)\frac{\partial \delta n_1}{\partial x}\right) \ee
where we have defined the two position dependent speeds of sound as
\be c^2_i(r) = \frac{g_{ii}n_i(r)}{m_i}\;\; i=1,2. \ee
The first component obviously behaves as it was alone and its dispersion law is for the 1D case \cite{menotti}:
\be \label{eq:dis1D}
\omega_{k,1}^2=\omega_{1}^2\frac{k(k+1)}{2}, \ee
where  $k$ depends on the mode one considers. We are interested in the center-of-mass (dipole) mode ($k=1$) and the lowest compressional (breathing) mode ($k=2$). Thus, for the majority component one has simply, $\omega=\omega_1$, for the dipole mode and $\omega=\sqrt{3}\omega_1$, for the breathing one.


For the second component an easy result is obtained by considering that for low frequency mode the spatial variation of the majority component is an higher order correction, i.e., $|\partial\delta n_2/\partial x| \gg |\partial\delta n_1/\partial x|$ and thus 
the left-hand side of Eq.\eqref{eq:hy2} can be neglected. In this case also the dispersion relation for the second component has the form Eq.\eqref{eq:dis1D} but where instead of the bare trapping frequency $\omega_2$  one has to use the renormalized frequency $\tilde{\omega}_2$ in Eq. (\ref{eq:myresult}). 
Such relation is valid for all the modes of the small condensate provided the excited modes are enough low frequency. It tells us that for all repulsive coupling constants the minority component dynamics is slower in presence of the majority one with respect of being alone. The result could have already been inferred from the analysis of the ground state in the previous section.

Notice that in this regime the minority component modes coincides essentially with the out-of phase mode and it is interesting the parameter $g_{22}$ does not enter at all in Eq. \eqref{eq:myresult}. This means that out-of-phase frequency can go to zero, but having nothing to do with the the phase separation. For homogeneous mixtures instead the softening (zero sound speed) of the out-of-phase mode -- which coincides in the highly unbalance case with the minority component mode -- is precisely the signature of phase separation \cite{PethickBook}.

Let us now study more accurately the collective mode frequencies by a direct simulation of the coupled GPEs Eqs. \eqref{eq:GPE1}-\eqref{eq:GPE2} and compare with the simple prediction Eq. \eqref{eq:myresult}. Details on the numerical method are given in the last session, but essentially we let Eqs. \eqref{eq:GPE1}-\eqref{eq:GPE2} evolve in time starting with the small component either squeezed or displaced with respect to the centre of the trap, depending on whether we want to study the breathing or the dipole mode, respectively.

In Fig. \ref{fig:freq-br} we show the results for the breathing mode of the minority component by varying the inter-species interaction strength $g_{12}$ and taking equal intra-species interaction strengths, $g_{11}=g_{22}$. One can see that the frequency of the minority component agree well with Eq.\eqref{eq:myresult} until $g_{12}/g_{11}\simeq0.8$, where one approaches the phase separation point, after which the mixed ground state here considered is unstable toward formation of domain walls as shown in the next section. The fact that the breathing mode starts increasing approaching the phase transition point is clear. Indeed, as shown in Fig. \ref{fig:DiffD}, the dimple in the bigger condensate becomes increasingly deep and the minority component starts being tightly trapped in it. This behaviour is in agreement to the recent analysis in \cite{jaksch}. Clearly the expression Eq. (\ref{eq:myresult}) derived above cannot be valid any more in this regime. 

Analogously in Fig. \ref{fig:freq-cm} we report the results for the center-of-mass mode frequency of the minority component.
Also in this case Eq.\eqref{eq:myresult} gives a pretty good estimate.
The frequency reduction close to the phase separation point could be interpreted, in this case, as an increase of the effective mass being the dimple moving together with the minority component, but this is not at all the whole story. 
Indeed the detailed analysis of the frequency enhancement/reduction beyond the phase transition point carried out in the next section, shows that we are dealing with a different object, whose dynamical behaviour strongly depends on its width. 

\section{Immiscible case}
\label{sec:imm}
\subsection{Ground state}

\begin{figure}[h*]
	\centering
	\includegraphics[width=7cm]{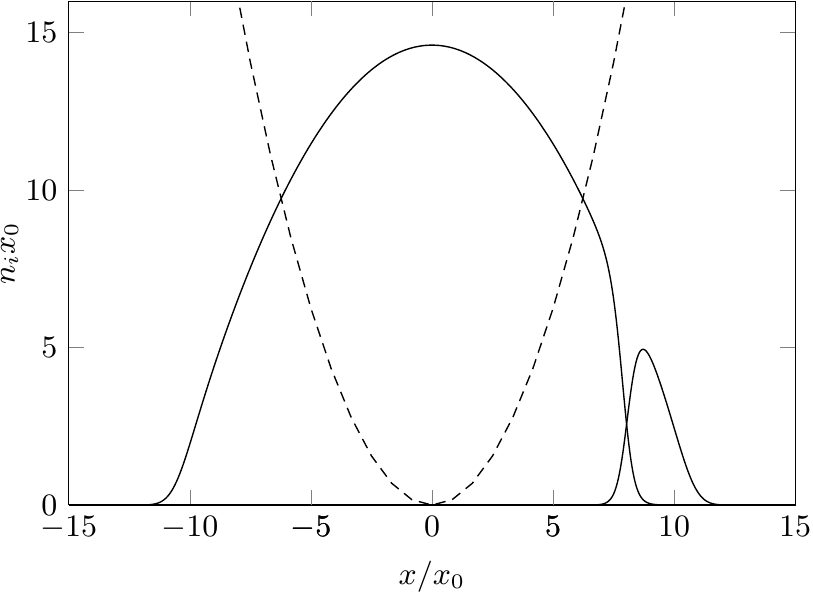}
\caption{Ground states for immiscible gases, i.e., $\Delta<1$: a typical configuration for equal trapping frequencies. The parameters are $\omega_1=\omega_2=2\pi\times200$~Hz and $g_{11}:g_{12}:g_{22}=1:1.5:1$, with $g_{11}=2.86 \times 10^{-37} $ Jm.} \label{fig:ps}
\end{figure}

\begin{figure}[h*]
	\centering
	\includegraphics[width=7cm]{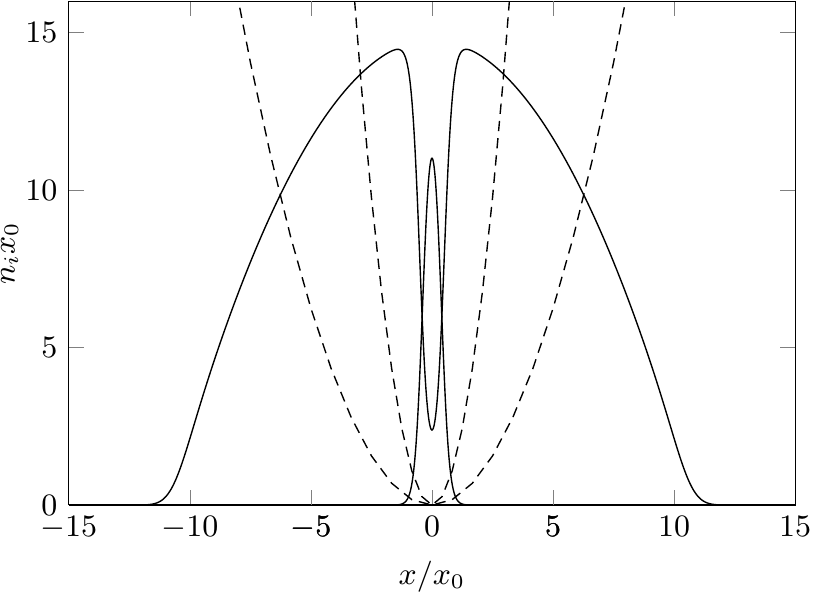}
	\caption{Ground states for immiscible gases, i.e., $\Delta<1$: a dDW soliton solution. The parameters are the same as for Fig. \ref{fig:ps}, but $\omega_2=2\pi\times500$~Hz.} \label{fig:gssol}
\end{figure}

\begin{figure}[t*]
	\centering
	\includegraphics[width=7cm]{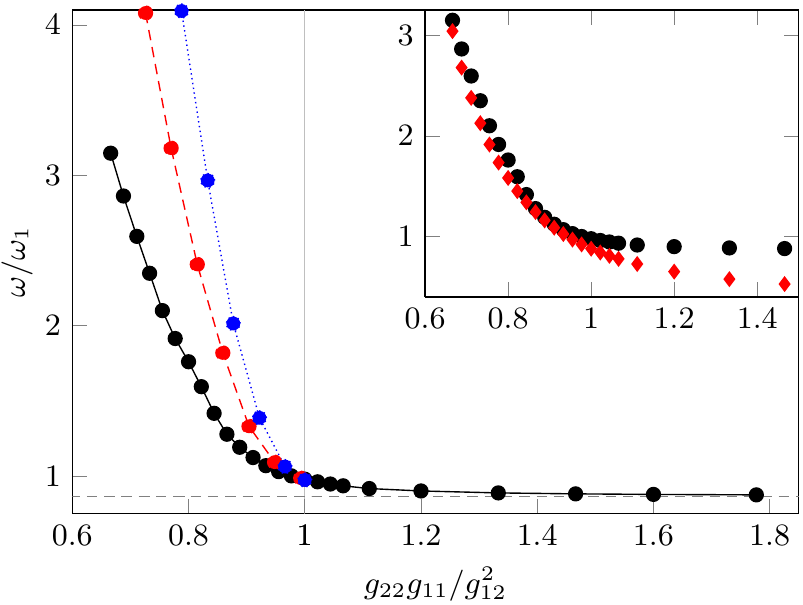}
	\caption{Breathing mode frequencies for the dDW soliton ground state (on the left of the vertical dotted line) as a function of $g_{22}$. For comparison we report again the results for the miscible case (right of the vertical dotted line). 
	The parameters are $N_1=200$, $N_2=10$, $m_1=m_2=87$~amu, $\omega_1/(2\pi)=\omega_2/(2\pi)=200$~Hz and $g_{11}=2g_{12}=2.86\times10^{-37}$~J$\cdot$m for the black points, $g_{11}=2g_{12}=4.27\times10^{-37}$~J$\cdot$m for the red ones and $g_{11}=2g_{12}=5.72\times10^{-37}$~J$\cdot$m for the blue ones. Dashed line represents the theoretical prediction \eqref{eq:myresult}. In the inset we plot the breathing mode frequency (black dots) together with the fitted parameter $1/(kd)$  (red diamonds). } \label{fig:Bsol}
\end{figure}

\begin{figure}[]
	\centering
	\includegraphics[width=7cm]{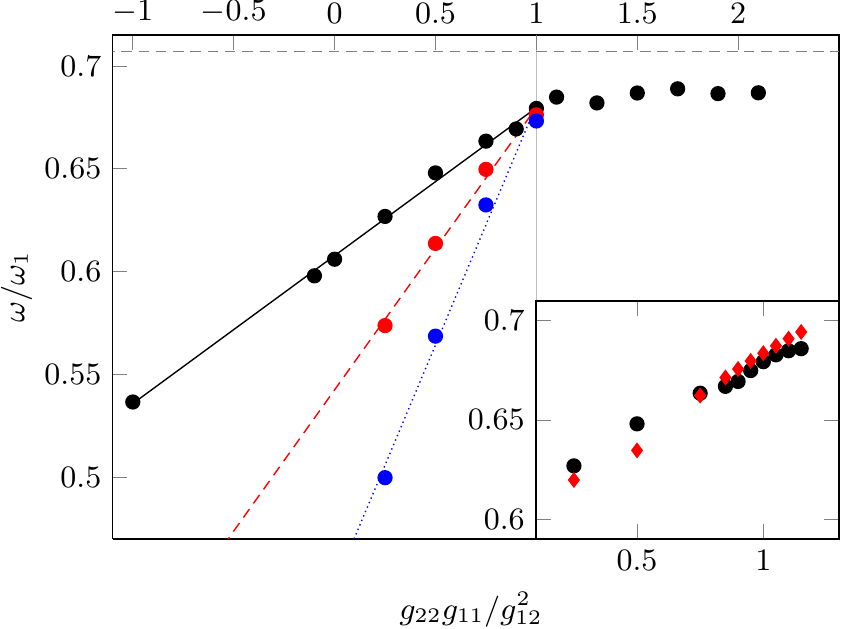}
	\caption{The same as Fig. \ref{fig:Bsol}, but for the centre of mass oscillation frequencies.  In the inset we plot the dipole mode frequency (black dots) together with the fitted funtcion $0.67+0.066 kd$  (red diamonds).}
\label{fig:CMsol}\end{figure}

So far we have considered only systems whose homogeneous ground state would correspond to a miscible mixture of the two gases. In this section we consider the case corresponding to an immiscible mixture, i.e., $\Delta<1$.
When this condition is satisfied the energy of the homogeneous phase becomes larger than the energy of a phase separated structure and the two components separate as can be seen in Fig. \ref{fig:ps}. Being a first order transition the structure present a domain wall (DW) between the two component.
Let us remind that in obtaining the phase separated density profiles with numerical methods one has to be a little careful. The ground states in Fig. \ref{fig:ps} have been obtained inserting some asymmetry in the initial wave functions, if they are perfectly symmetric one finds the wrong ground state in which the smaller component remains at the centre of the bigger one or it splits up in two parts located on the sides (both these solutions have a bigger energy). A recent and detailed discussion about this issue can be found in \cite{pattinson}. 

Another and more interesting ground state is obtained when the trapping frequency of the minority component is bigger than the one of the majority component. As shown in Fig.~\ref{fig:ps}, a ground state with the minority component at the centre of the trap is possible\cite{MixturesSvid}. Indeed, in this case,  the energy gained by separation is less than the energy lost by going in a region in which the trapping potential is higher. Although the density profiles do not show any drastic change by going across the phase transition point (see Fig.~\ref{fig:DiffD}), the states differ a lot concerning their dynamical properties. The point is that the gas is in the immiscible phase and the new structure is made out of two domain walls connected by the quantum pressure. As we show in the next section such structure has a solitonic nature. In order to distinguish it from other solitonic structures in Bose-Bose mixtures (see, e.g., \cite{Families} and reference therein) and from other domain wall cases (see e.g. \cite{sDWsol}), we call it double domain wall (dDW) soliton.
Here, we use the term ``soliton" for a non-dispersive localized structure stable against collision in agreement with the definitions in \cite{soliton}.

\subsection{Collective modes and collisions}

As for the miscible case we start studying the dynamics of the breathing mode. The frequency is obtained by numerically evolving Eqs.\eqref{eq:GPE1}-\eqref{eq:GPE2} starting with a dDW structure and suddenly changing the trapping frequency of the minority component  to the value of the majority component one. 
The results are reported in Fig. \ref{fig:Bsol}. For comparison we show again also the miscible case results. With respect to the latter case, one immediately see that in the immiscible case the frequency dependence on the interaction strength parameters $g_{ij}$ is very different and not universal. 
However the general trend is quite natural, i.e., the narrower the density hole, the larger is the frequency.
Indeed the frequency increases both decreasing $g_{22}$ and/or by increasing $g_{11}$. 
This is also consistent, but less intuitive, with the behaviour of the dipole mode frequency, reported in Fig. \ref{fig:CMsol}. In this case the narrower the hole the smaller the frequency.

In order to better understand such trend we fit the soliton solutions $\Psi_{dDW}$ with a simple double domain wall ansatz in analogy with the single DW profile (see, e.g., \cite{Families})
\be
\Psi_{dDW}(x)\propto \tanh(k(x+d)) - \tanh(k(x-d)),
\ee
with $k$, $d$ positive real numbers. As it can been seen from the insets of Figs. \ref{fig:Bsol} and  \ref{fig:CMsol} 
the mode frequencies are, as expected, strictly related to the adimensional parameter $k d$, which is a sort of effective width of the dDW. In particular we find that in the solitonic regime $k d\propto \Delta$ and the breathing mode frequency is inversely proportional to $g_{22}$ or equivalently to $k d$ at fixed $g_{11}$ and $g_{12}$, while the dependence of the dipole mode frequency is linear (Figs. \ref{fig:Bsol}-\ref{fig:CMsol}).

\begin{figure}[t*]
	\centering
	\includegraphics[width=7cm]{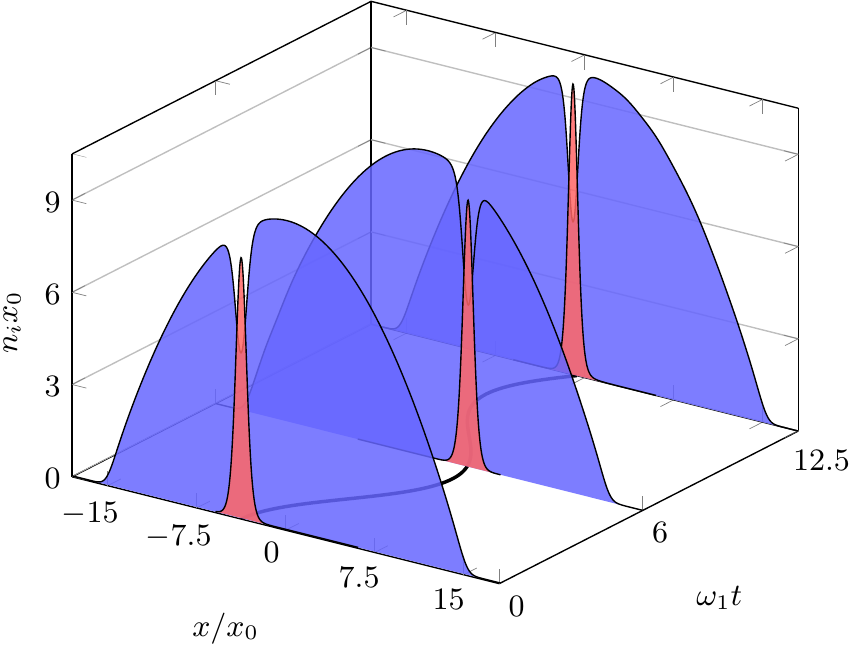}
	\caption{Time evolution of a displaced dDW soliton. }
\label{fig:dipoleDyn}\end{figure}

As we already anticipated the dDW behaves like a soliton. The first evidence is its dynamics inside the trap, shown in Fig. \ref{fig:dipoleDyn}. The dDW structure remains localized and its motion is not dispersive and it is globally well defined after many periods of oscillation. Viceversa in the miscible case the motion of the impurity is dispersive and it excites a lot of phonons in the majority component already in the first period of the oscillation. The oscillation itself indeed loses to be well defined after a few periods. 
  
\begin{figure}[h*]
	\centering
	\includegraphics{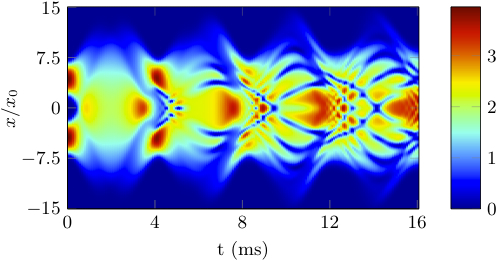}
	\caption{Time evolution of two colliding impurities in the miscible $\Delta>1$ regime.Already after the first collision the impurities lose their identity.} 
\label{fig:nosolmap}
\end{figure}

\begin{figure}[h*]
	\centering
	\includegraphics{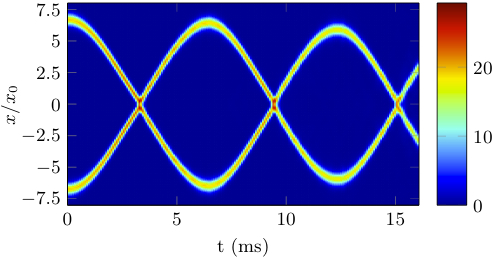}
	\caption{Time evolution of two colliding solitonic domain walls. The localized, non-dispersive motion is evident and they preserve they nature after many collisions.} \label{fig:solmap}
\end{figure}
Finally we address the behaviour of the dDW against collisions with respect to the the miscible case. For this we simulate a collision between two very small cloud of minority component. We find that the collisional behaviour is even more different depending on whether the gases are miscible or not as clearly shown in Fig. \ref{fig:nosolmap} and Fig. \ref{fig:solmap}, respectively. 
In the earlier case, $\Delta>1$, already after the first collision, dispersion and interference appear and the two impurities lose soon their identity. On the contrary the dDW solitons ($\Delta<1$) keep being well defined objects even after many collisions, justifying (within our definition) the name soliton for such a structure.

\section{Conclusions}

In conclusion we show that dynamics of the minority component of an highly unbalanced Bose-Bose mixture can be pretty reach already at the mean-filed level.

In particular in the miscible phase we provide analytical expressions for the frequency modes, Eq. \ref{eq:myresult}.
Such prediction is confirmed by a numerical analysis based on coupled time-dependent Gross-Pitaevskii equations.

In the immiscible phase we find an interesting ground state solution which is localized, non-dispersive and stable against collisions. Thanks to such properties we name it a double domain wall (dDW) soliton.
The dynamics of such structure is studied in details numerically and an interesting dependence on the interaction strengths has been found. 

Although it could be clear it is worth reminding that the dDW soliton differ from of dark-bright soliton \cite{busch}. The latter is an excited state of the system with a peculiar phase, which exists for $\Delta>1$. Moreover by removing the bright component the dark one is still a metastable state of the gas.
The dDW soliton instead exists only for $\Delta<1$, it is a ground state of the Bose-Bose mixture and is unstable against removing the (interior) minority component.

It is also worth mentioniong that has been recently argued the dDW soliton configuration (although not call in this way and for other purpose) can show interesting density correlations \cite{busch-corr}.

The measurement of the frequency modes in the various regime discuss in the present Letter could also be a way  to precisely determine the coupling constants $g_{ij}$. In particular the oscillation of the dDW soliton should be long living, with respect to the miscible case, thus allowing for a long time period measurement. Moreover since in general the one-dimensional system is realized by strongly confining the gases in two directions $g_{ij}$ is directly related to the three dimensional $s$-wave scattering lengths $a_{ij}$.
A similar strategy has been pursued in a recent paper by Egorov \emph{et al.} \cite{ergov} for a cigar-shaped trap in the miscible case. They indeed found that the collective oscillation is determined by the parameter $a_{12}/a_{11}$ and, fitting numerical simulation results with experimental ones, they obtained a value of $a_{12}$ for a particular mixture in a reasonable agreement with theoretical prediction.

The results obtained in this work are also interesting in view of recent experiments in the direction of spin dynamics. Let us first of all mention the one realized by Catani \emph{et al.} \cite{catani}. They studied the breathing oscillation of a potassium impurity in a one dimensional tube filled by rubidium atoms at different values of the rubidium-potassium interaction. In this experiment the temperature is at the moment too high to see the effect studied here. 

Very low temperature has been instead reach in the presence of an optical lattice in the experiments performed by Fukuhara \emph{et al.} \cite{impBECBloch}. In particular they already studied the motion of an impurity in the polaronic regime. The presence of a lattice as long as the bath is weakly interacting (well in the superfluid regime) would not alter qualitatively the physics we describe in this work.

\section{Numerical method}\label{numerics}
We used a modified version of the Fourth-order Runge-Kutta in the Interaction Picture (RK4IP) algorithm developed by Ballagh \emph{et al.} \cite{ballagh} (and appendix A of \cite{norrie}) to propagate the stochastic differential equations obtained with the truncated Wigner method \cite{sinatra}.

The oscillation frequencies are obtained in two steps. First we find the proper ground state of the mixture by imaginary time evolution and letting all the high energy component of the initial guess wave-function to die. For the breathing mode the ground state is determined in the presence of two trapping harmonic potentials centered in zero, where the frequency for the smaller component is much larger than for the majority one. If instead we are interested in the CM oscillations the potential for the second component is also centered in a point different from zero. After we got the wave-functions, these are used as input of the program in real time that evolves it for equal (species independent) trapping potentials. The mode frequencies are determined by fitting the mean square root radius for the breathing and the mean position for the dipole one.

\acknowledgments

We would like to thank Marta Abad for useful suggestions and for the reading of the present manuscript. 
We also acknowledge useful discussions with Franco Dalfovo, Giacomo Lamporesi, Lev Pitaevskii and Sandro Stringari. This work has been supported by ERC through the QGBE grant and by Provincia Autonoma di Trento.

\end{document}